\documentclass{aa}

\usepackage{graphicx}

\usepackage{txfonts}

\usepackage{amssymb}
\usepackage{amsmath}
\usepackage{graphicx}
\usepackage[colorlinks=true, allcolors=blue]{hyperref}
\usepackage{booktabs}
\usepackage{amsfonts}     
\usepackage[T1]{fontenc} 
\usepackage[utf8]{inputenc} 
\usepackage{multicol}
\usepackage{xcolor}
\usepackage[switch]{lineno}

\usepackage{soul}
\usepackage{rotating}
\usepackage{comment}
\usepackage{multirow}
\usepackage{longtable}
\usepackage{lscape}
\usepackage{color}
\usepackage{longtable}
\usepackage{url}

\usepackage{academicons}

\begin{document}

\title{Optical polarization variability and its relation to $\gamma$-ray activity in blazars}

\author{
Styliana Grigoriou\inst{\ref{AstroCrete2}} \and
Anastasia Glykopoulou\inst{\ref{AstroCrete},\ref{AstroCrete2}} \and
Ioannis Liodakis, \inst{\ref{AstroCrete}, \ref{AstroBioCrete}} \thanks{\href{mailto:liodakis@ia.forth.gr}{liodakis@ia.forth.gr}} \orcid{0000-0001-9200-4006} 
}

\institute{
Department of Physics, University of Crete, GR-70013 Heraklion, Greece \label{AstroCrete2} 
\and
Institute of Astrophysics, Foundation for Research and Technology-Hellas, GR-70013 Heraklion, Greece \label{AstroCrete} 
\and
Institute of Astrobiology, University of Crete, GR-70013 Heraklion, Greece \label{AstroBioCrete} 
}

\abstract{Optical polarization can be an important probe of particle acceleration and high-energy emission processes in relativistic jets from supermassive black holes. We combined publicly available observations from three past blazar monitoring programs to produce densely sampled light curves for 15 blazars in order to explore the relation of $\gamma$-ray activity to the polarization variability as well as discover new rotations of the polarization angle. We find that the polarization degree does not correlate with the $\gamma$-ray flux for individual sources nor different subsamples of blazars, potentially indicating multiple emission mechanisms. In the combined dataset, we identified a total of 64 rotations in 12 sources, 39 of which are newly identified rotations.  We confirm the trend found in previous works for the whole sample: lower polarization degrees during periods of polarization angle rotations. However, looking at the individual sources, we identified cases where the rotation and non-rotation polarization degree distributions are indistinguishable, providing further evidence for the multiple emission mechanism hypothesis.}

\keywords{Polarization -- Relativistic processes -- Galaxies: active -- BL Lacertae objects: general -- Galaxies: jets}

\titlerunning{Polarization variability and $\gamma$-ray activity}
\maketitle
\nolinenumbers 

\section{Introduction}

Active galactic nuclei with powerful jets are often the brightest objects in the Universe. The very brightest active galactic nuclei are blazar, whose jet is aligned with our line of sight \citep{Blandford2019,Hovatta2019}. Blazars are known for their bright outbursts and fast variability across the electromagnetic spectrum. They are also known for their highly polarized emission arising from synchrotron radiation from relativistic electrons in the jet.  Optical polarization is an important probe of the magnetic field structure and evolution in the jets; however, the stochastic nature of the variability requires long-term monitoring programs that are very demanding in terms of both telescope time and people-power. Nevertheless, a lot of effort has been focused on understanding the underlying drivers of polarization variability, including several past and current monitoring programs \cite[e.g.,][]{Bachev2012,Agudo2012,Jermak2016,Uemura2017,Blinov2021,Raiteri2021-II,Marscher2022,Polychronakis2025}.

One such program was the RoboPol survey \citep{Blinov2021}. RoboPol observed a large number of blazars with a regular cadence of a few days over four years. This allowed us to examine the variability statistics of the population and uncover hidden aspects of jet physics. We also found that, at least in a few sources, polarization can vary on shorter timescales than the nominal few-day cadence of the RoboPol survey \citep{Kiehlmann2021}. Motivated by these results, we used publicly available data from three major blazar monitoring programs to build the best-sampled polarization light curves for 15 common sources. Our objective was twofold: (i) explore the relation between polarization degree and $\gamma$-ray activity and (ii) uncover new rotations of the polarization angle that  either  were missed due to the limited cadence of the individual monitoring programs or were unreported. Our paper is organized as follows. In Sect. \ref{sec:data} we describe how we built the polarization light curves, in Sect. \ref{sec:res} we explore the connection between the polarization degree and $\gamma$-ray activity and report on newly found rotations, and in Sect. \ref{sec:disc} we discuss our findings and present our conclusions.

\section{Data and analysis} \label{sec:data}

\begin{figure*}
\centering
\includegraphics[width=1\textwidth]{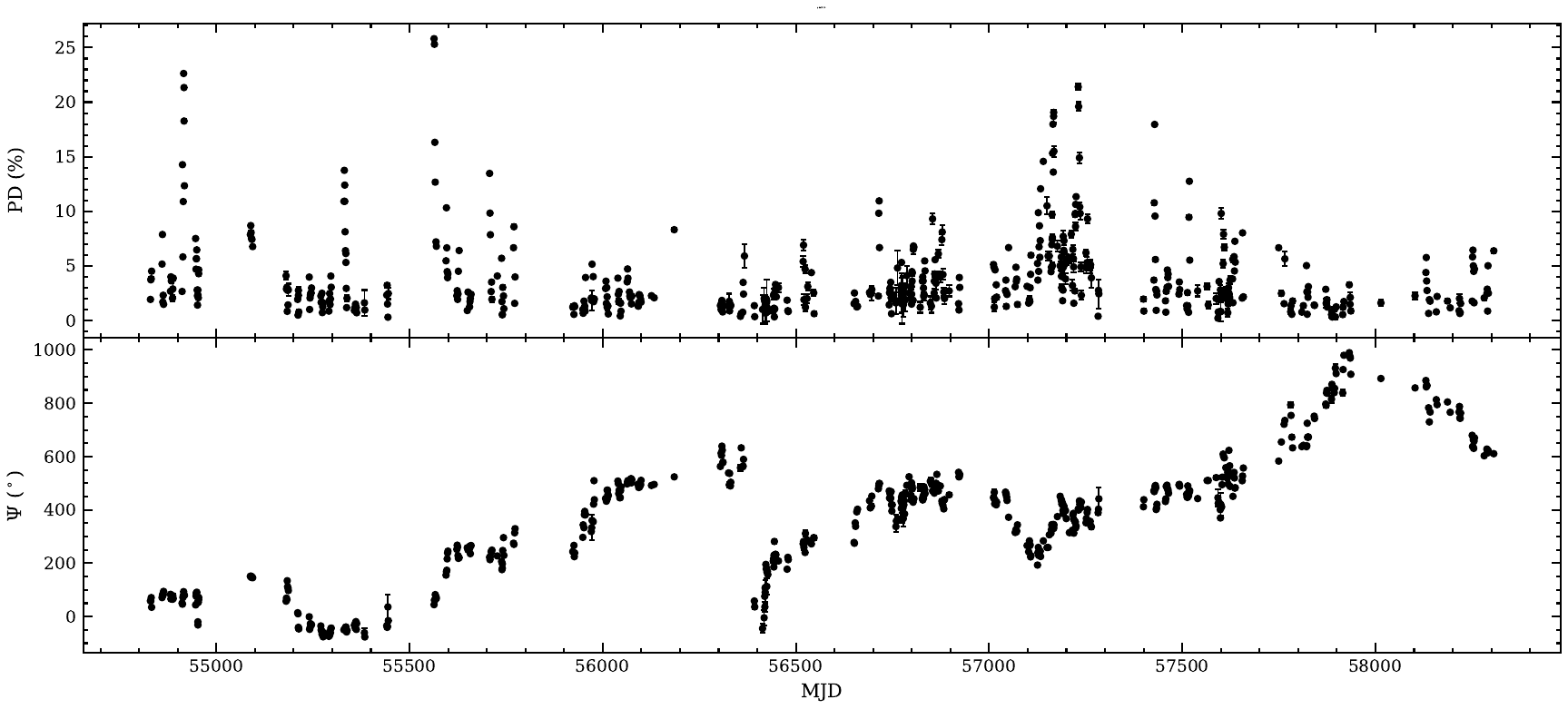}
\caption{Optical polarization degree (top) and angle (bottom) light curve for PKS~1510-089. The polarization angle has been corrected for the 180$^\circ$ ambiguity.}
\label{plt:pks1510_lightcurve}
\end{figure*}

We used publicly available data from three past major blazar monitoring programs, namely RoboPol \citep{Blinov2021},  Kanata \citep{Itoh2016}, and the Steward Observatory \citep{Smith2009}. We focused on the polarization degree and polarization angle in the R band, which was covered by all three programs.  After collecting all available data from each telescope and observatory, we cross-matched the different source names used by each facility in order to identify common blazars. We identified 20 such sources; however, upon visual inspection of the light curves, we excluded five sources due to insufficient coverage. The final sample of 15 sources is presented in Table~\ref{tab:crossmatch}. An example of a light curve is shown in Fig. \ref{plt:pks1510_lightcurve}, for PKS~1510-089. The combined data as provided by the respective observatories are available in the Harvard Dataverse repository\footnote{\url{https://dataverse.harvard.edu/dataset.xhtml?persistentId=doi:10.7910/DVN/5UIUED}} \citep{Grigorioudata2026}.

For all the sources, we used publicly available $\gamma$-ray light curves from the \textit{Fermi} Gamma-ray Space Telescope light curve repository \citep{repository2023}\footnote{\url{https://fermi.gsfc.nasa.gov/ssc/data/access/lat/LightCurveRepository/}}. We extracted the energy flux, 3-day-binned light curves in the 0.1--100~GeV range  with a test statistic threshold of four and the spectral index as a free parameter.

To identify periods of significant variability in the $\gamma$-ray and polarization degree light curves, we used Bayesian blocks (BB; \citealp{Scargle2013}). BB has only one parameter, which was set to $p_0 = 0.01$, i.e., a 1\% false-positive detection of a new block. We applied BB to the polarization degree light curves and then mapped the times of the blocks to the $\gamma$-ray light curve. That way we could determine whether a change in the polarization degree produces a response in the $\gamma$-rays. We then computed the weighted average of the blocks for both the polarization degree and $\gamma$-rays in order to separate ``active'' and ``non-active'' states. 

For the polarization angle, we used the automatic detection algorithm for polarization angle rotations detailed in \cite{Glykopoulou2026}. In summary, the algorithm uses BB to separate the polarization angle light curve into periods of continuous change (either positive or negative). Sections of the light curve with an amplitude of $>90$ degrees are then tested using the student t-test and binomial test. P-values lower than 0.05 and 0.065 respectively indicate the presence of a polarization angle rotation.

 \section{Results} \label{sec:res}

\subsection{Polarization degree variability}
\begin{figure}
\centering
\includegraphics[width=0.45\textwidth]{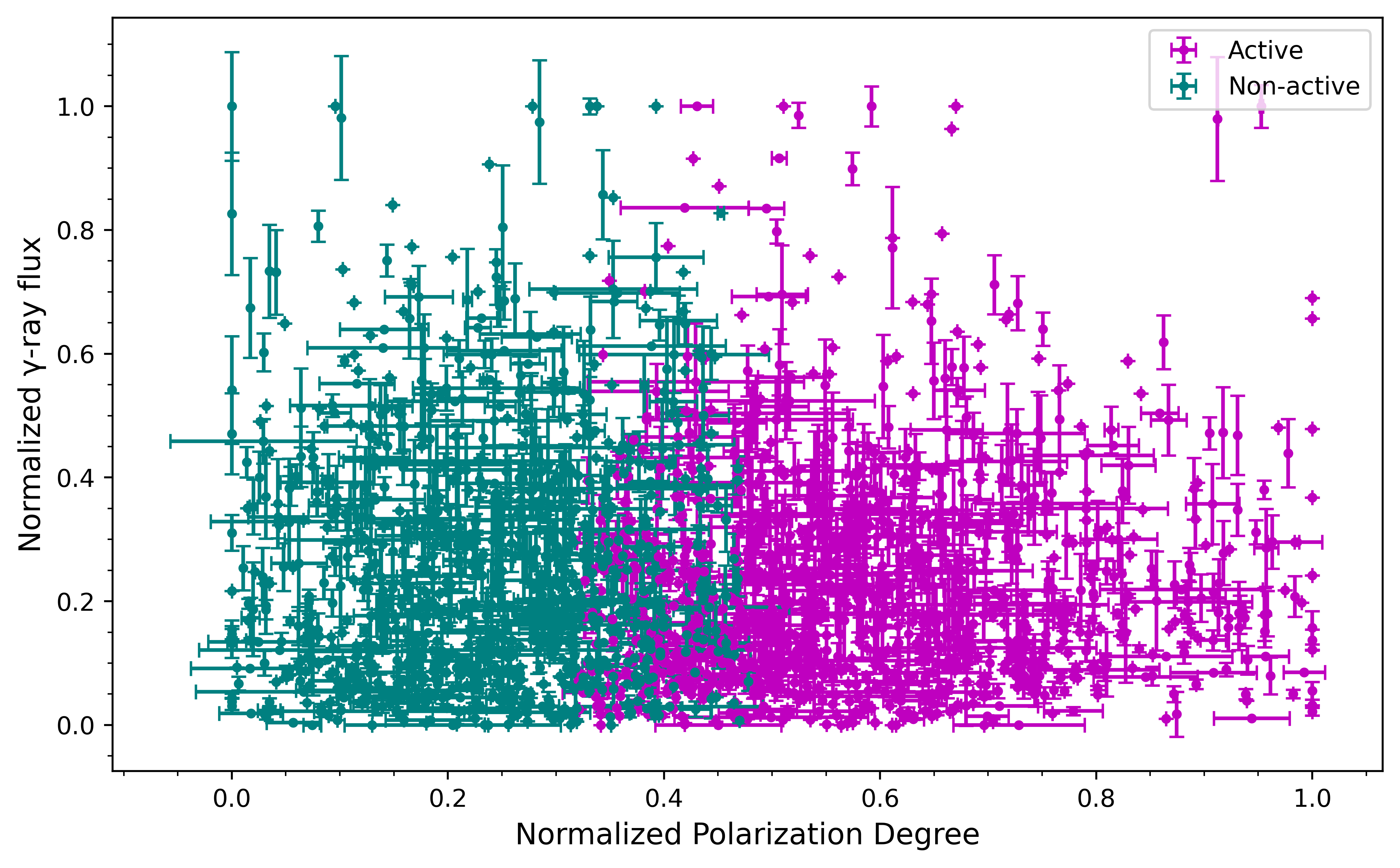}
\caption{Normalized $\gamma$-ray flux versus normalized polarization degree for BLLs. The colors denote the active and quiescent polarization states defined as above and below the median polarization degree. The Spearman results support neither a correlation nor an anticorrelation for the quiescent, active, or combined datasets.}
\label{plt:gamma_vs_pol}
\end{figure}

We used the results of the BB and the Spearman test to investigate potential correlations between the individual sources as well as the different subclasses of blazars. We considered the two sub-classifications, flat spectrum radio quasars (FSRQs) and BL Lac-type objects (BLLs),  as well as the low-, intermediate-, and high-synchrotron peak (LSP, ISP, and HSP, respectively) blazars (\citealp{Ajello2020}). The final sample contained 15 objects: 11 BLLs and 4 FSRQs. In terms of synchrotron-peak classification, the sample includes 7 LSP, 4 ISP, and 4 HSP blazars. To combine difference sources, we first normalized both the polarization degree and $\gamma$-ray flux by their respective medians to remove any offset from intrinsically more polarized or more $\gamma$-ray-bright sources. Examples of the polarization degree versus $\gamma$-ray flux are shown in Fig. \ref{plt:gamma_vs_pol}. The results of the Spearman correlation analysis that included $r_s$ and associated p-values are presented in Tables \ref{tab:spearman_results} and \ref{tab:spearman_groups}.

Overall, no strong correlation between the polarization degree and $\gamma$-ray flux was observed. There are hints of an anticorrelation in different blazars; however, they are typically not statistically significant. Two sources show a mildly significant anticorrelation according to the Spearman test. Those were 3C~66A in the non-active state and 3C~454.3 considering all the observations. However, visual inspection of the correlation does not provide sufficient evidence. It is likely that in the first case the anticorrelation is driven by a few outliers, and in the latter case it is driven by the  combination of the active and non-active states, while none of the states shows  a clear correlation. Therefore, we considered both to be artifacts and not real anticorrelations. When sources are grouped by type, all of the subsamples appear uncorrelated across all activity states. However, we caution the reader that the low number of sources in the different subclasses prevents us from drawing strong conclusions.

\subsection{Polarization angle variability}
\begin{figure}
\centering
\includegraphics[width=0.45\textwidth]{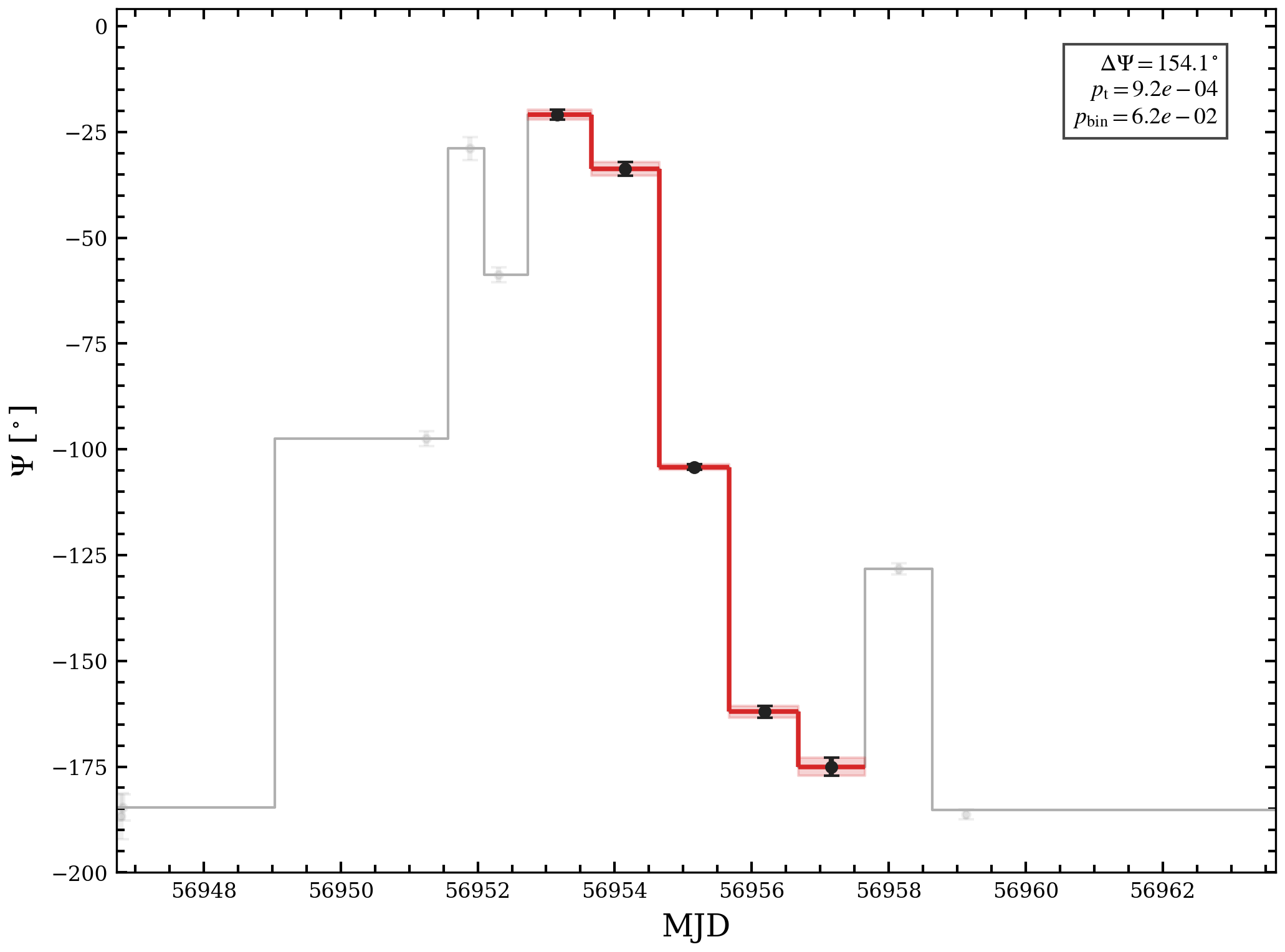}
\caption{Optical polarization angle rotation of about 154 degrees in CTA 102. The red line shows the identified rotation. The legend lists the p-values for the t-test and binomial tests. }
\label{plt:CTA102_rotation}
\end{figure}

\begin{figure}
\centering
\includegraphics[width=0.45\textwidth]{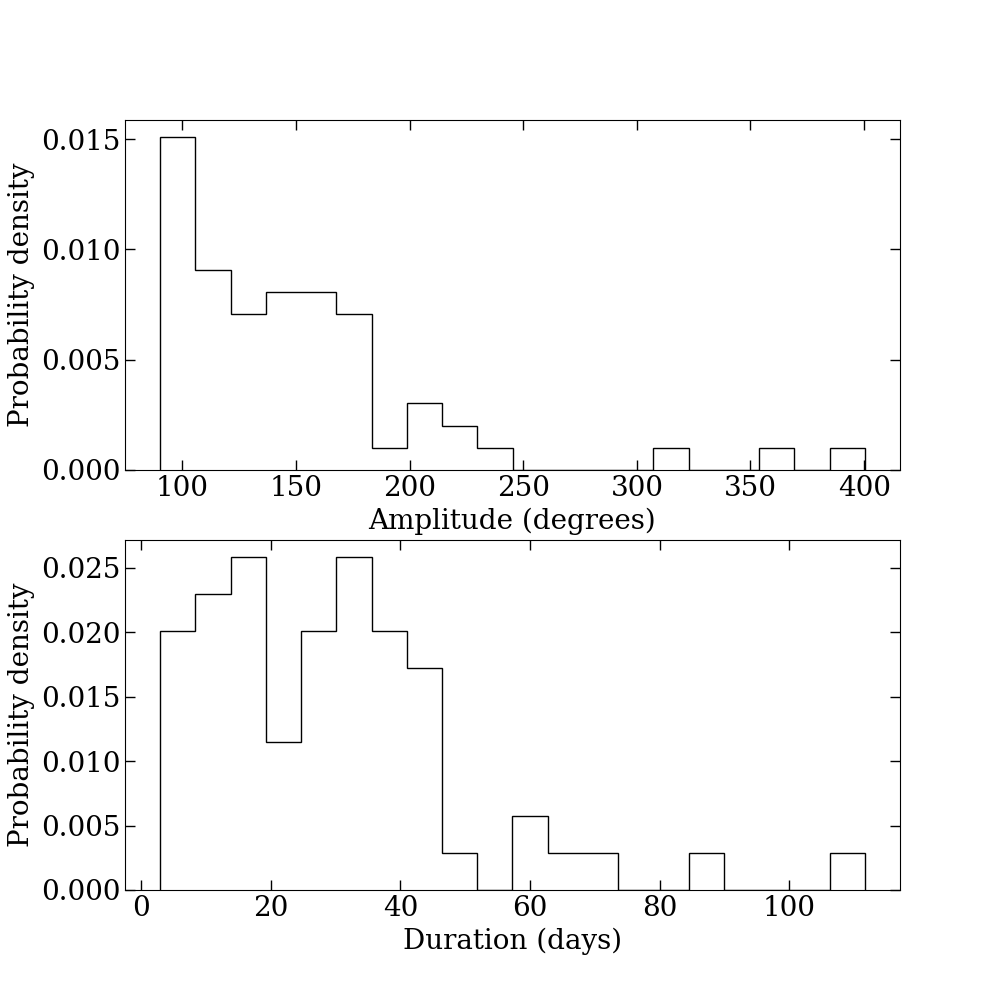}
\caption{Distribution of amplitudes (top) and duration (bottom) for the rotations in our sample.}
\label{plt:ampl_dur}
\end{figure}

\begin{figure}
\centering
\includegraphics[width=0.45\textwidth]{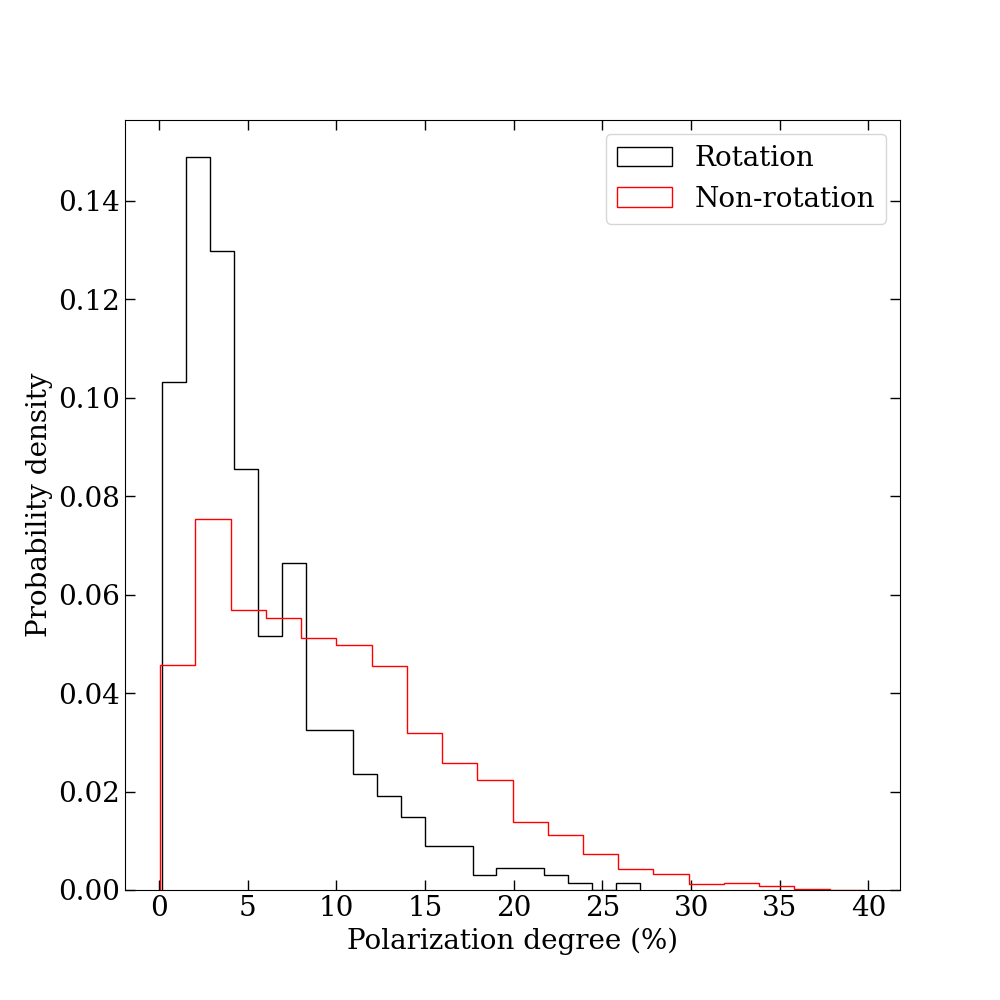}
\caption{Distribution of the polarization degree during the rotation and non-rotation periods of all the sources in our sample combined.}
\label{plt:rot_nonrot_pd}
\end{figure}

Using the polarization angle light curves and the methodology outlined in \cite{Glykopoulou2026}, we identified 64 rotations in 12 sources. Figure \ref{plt:CTA102_rotation} shows an example of a polarization angle rotation, and Fig. \ref{plt:ampl_dur} shows the distribution for the amplitude and duration of the identified rotations. We find an average amplitude of 137 degrees and an average duration of 28 days. The maximum  amplitude is 400 degrees, and the maximum duration is 111 days, which happened in AO~0235+164 and PKS~1510-089, respectively. All the rotations are listed in Table \ref{tab:rotations}. Three sources did not show any rotations: Mrk~501, 1ES 1959+650, and 1ES 2344+514. Unsurprisingly, these three are all HSP blazars; as demonstrated in \cite{Blinov2016-II}, HSP blazars are less likely to produce rotations. 

Of the 64 rotations, 21 have already been reported in \cite{Blinov2015,Blinov2016-II,Blinov2018} and \cite{Glykopoulou2026}. We also found four more previously reported in the literature (see Table 3 in \citealt{Blinov2015} and references therein). This leaves us with 39 newly discovered rotations. However, we note that our search is not extensive; some rotations may have been included in publications that we missed. 

Given the length of the light curves and the large number of rotations, we investigated the rotation relationship to the polarization degree. For each source with identified rotations, we separated the polarization degree light curve into periods of rotation (i.e., the polarization degree within the start and end date of the rotation) and periods outside of a rotation, which we then compared using the Anderson-Darling test. Figure \ref{plt:rot_nonrot_pd} shows the comparison of the polarization degree for the rotation and non-rotation periods for all the sources in the sample. We find that for individual sources there is an overall preference to reject the null hypothesis that the two distributions are coming from the same parent population. There are three exceptions, namely AO~0235+164, PKS~1749+096, 3C~454.3. Looking at the whole sample and considering the median polarization degree for each source, we marginally (4.5\%) reject the null hypothesis and confirm the result of \cite{Blinov2016}, that the polarization degree tends to be lower during periods of rotation.

\section{Discussion and conclusions}\label{sec:disc}

In this work we used publicly available data to produce densely sampled light curves for a small number of blazars. Our objective was twofold: (i) to investigate the relation between the polarization degree and $\gamma$-ray activity during active and non-active states and (ii) to discover new rotations of the polarization angle that could have been missed because of incomplete coverage. 

Comparing the polarization degree to the $\gamma$-ray flux revealed no strong correlation in either individual sources or subclasses of blazars. We tested for simultaneous variations between the polarization degree and $\gamma$-rays. This is expected since the co-spatiality of the optical and $\gamma$-ray emission regions has been demonstrated  through several cross-correlation analyses \citep{Liodakis2018,Liodakis2019,deJaeger2023}. Follow-up analysis that takes time delays into consideration could provide interesting results. We also note that different underlying $\gamma$-ray emission processes --- external Compton (EC) versus synchrotron self-Compton (SSC) --- could be operating in different sources of the same blazar subclass \citep{Liodakis2019}, and potentially even in the same source at different times. Due to the different response of the $\gamma$-ray emission to variations in the seed photon field in EC and SSC scenarios, using median quantities and/or the entire light curve might not fully capture the physically distinct behaviors in different blazar subclasses. This additional complexity could potentially introduce biases that dilute any intrinsic correlation. This can be addressed by identifying flares in individual sources through dedicated flare modeling,  combined with detailed spectral energy distribution modeling to isolate SSC from EC flares. We reserve this analysis for a future publication.

Exploring the polarization angle light curves, we found 64 rotations in 12 sources, 39 of which had not been previously reported to our knowledge.  There is a wide variety of amplitudes and durations, with an average of 137 degrees and 28 days, respectively. All of the LSP and ISP blazars show rotations, whereas only one out of the four HSP blazars does. This is in line with previous RoboPol results that HSP blazars rotate less frequently \citep{Blinov2016-II} as well as the emerging energy-stratified picture \cite[e.g.,][]{Angelakis2016,Tavecchio2018,Tavecchio2021,Liodakis2022}, which suggests that HSP blazars rotate in X-rays. HSP polarization angle rotations in X-rays have in fact been recently found by the Imaging X-ray Polarimetry Explorer  \citep{Soffitta2023,DiGesu2023,Maksym2025}.

We also investigated the behavior of the polarization degree during periods of rotation, which we compared to periods without rotation. We confirm the trend identified by RoboPol of lower polarization degrees during rotations for the whole sample. However, we have also identified cases where we cannot reject the null hypothesis that the two distributions are coming from the same population, i.e., there is no difference in the polarization degree. The drop in polarization degree is expected in both plasmoid-driven magnetic reconnection models \cite[e.g.,][]{Zhang2018,Hosking2020,Zhang2020} and models involving two polarization components with a different polarization angles and a varying contribution to the total flux  \cite[e.g.,][]{Cohen2020,Liodakis2020}. There are now examples of an increase in the polarization degree during $\gamma$-ray flares that can be attributed to changes in the Doppler factor of an emission component, likely a shock propagating in the jet  \cite[e.g.,][]{Marscher2010,Jorstad2013,Raiteri2017,Liodakis2020}. Overall, our results do not favor one mechanism over another and instead point to multiple emission mechanisms operating at different times in the same blazars. However, we do caution the reader that our sample is limited. Larger samples are necessary to confirm our results. 

\section*{Data availability}
The combined multi-telescope photometry and polarimetry data presented in this work are available in the Harvard Dataverse repository \url{https://dataverse.harvard.edu/dataset.xhtml?persistentId=doi:10.7910/DVN/5UIUED} and in electronic form at the CDS via anonymous ftp to cdsarc.u-strasbg.fr (130.79.128.5) or via \url{http://cdsarc.u-strasbg.fr/viz-bin/cat?J/A+A/2026/A78}.
\\

\begin{acknowledgements}
We thank the anonymous referee for comments that helped improve this work. We also thank Carolina Casadio and Vasiliki Pavlidou for discussions that helped improve this work. AG and IL were funded by the European Union ERC-2022-STG - BOOTES - 101076343. Views and opinions expressed are however those of the author(s) only and do not necessarily reflect those of the European Union or the European Research Council Executive Agency. Neither the European Union nor the granting authority can be held responsible for them. Data from the Steward Observatory spectropolarimetric monitoring project were used. This program is supported by Fermi Guest Investigator grants NNX08AW56G, NNX09AU10G, NNX12AO93G, and NNX15AU81G. This research was partially funded by the Deutsche Forschungsgemeinschaft (DFG, German Research Foundation) as part of the DFG Research Unit FOR5195 – project number 443220636.
\end{acknowledgements}

\bibliographystyle{aa}  
\bibliography{bibliography}

\begin{appendix}

\section{Blazars detected  by all three monitoring programs}
\begin{table*}
\centering
\caption{Cross-matching source identifiers across telescopes.}
\begin{tabular}{cccc}
    \hline
    Robopol & Steward  &  Kanata & Fermi  \\
    \hline
    RBPLJ0222+4302 & 3C 66A & Kanata-05 & 4FGL J0222.6+4302 \\
    
    RBPLJ0238+1636 & AO 0235+164 & Kanata-06 & 4FGL J0238.6+1637\\
    
     RBPLJ0721+7120 & S5 0716+714 & Kanata-12 & 4FGL J0721.9+7120\\
    
    RBPLJ0738+1742 & OJ 287 & Kanata-18 & 4FGL J0854.8+2006\\
   
   RBPLJ1221+2813 & W Com & Kanata-24 & 4FGL J1221.5+2814\\
    
    RBPLJ1256-0547 & 3C 279 & Kanata-27 & 4FGL J1256.1-0547\\
    
    RBPLJ1512-0905 & PKS 1510-089 & Kanata-30 & 4FGL J1512.8-0906\\
    
    RBPLJ1555+1111 & PG 1553+113 & Kanata-32 & 4FGL J1555.7+1111\\
    
    RBPLJ1653+3945 & Mrk 501 & Kanata-34 & 4FGL J1653.8+3945\\
    
    RBPLJ1751+0939 & PKS 1749+096 & Kanata-37 & 4FGL J1751.5+0938\\
    
    RBPLJ1959+6508 & 1ES 1959+650 & Kanata-40 & 4FGL J2000.0+6508\\
    
    RBPLJ2202+4216 & BL Lac & Kanata-42 & 4FGL J2202.7+4216\\
    
    RBPLJ2232+1143 & CTA 102 & Kanata-43 & 4FGL J2232.6+1143\\
    
    RBPLJ2253+1608 & 3C 454.3 & Kanata-44 & 4FGL J2253.9+1609\\
    
    RBPLJ2347+5142 & 1ES 2344+514 & Kanata-45 & 4FGL J2347.0+5141\\
    \hline   
\end{tabular}
\label{tab:crossmatch}
\end{table*}

\section{Results from the Spearman correlation analysis}
\begin{table*}
\centering
\caption{Spearman correlation results for individual blazars.}
\begin{tabular}{lcccccc}
\hline
Name (Type)& \multicolumn{2}{c}{Active states} & \multicolumn{2}{c}{Non-active states} & \multicolumn{2}{c}{All states} \\
\hline
 &  $r$ & $p$ & $r$ & $p$ & $r$ & $p$ \\
\hline
3C 66A (BLL,ISP) & -0.15 & 4.91e-02 & -0.38 & 6.20e-04 & -0.20 & 1.38e-03 \\

AO 0235+164 (BLL,LSP) & 0.13 & 4.76e-01 & -0.08 & 4.19e-01 & 0.03 & 7.35e-01 \\

S5 0716+714 (BLL,ISP) & 0.20 & 4.22e-02 & -0.05 & 6.22e-01 & -0.02 & 7.41e-01 \\

OJ 287 (BLL,LSP) & 0.15 & 5.71e-02 & 0.05 & 6.06e-01 & 0.21 & 3.69e-04 \\

W Com (BLL,ISP) & 0.04 & 6.68e-01 & 0.17 & 1.82e-01 & -0.14 & 5.56e-02 \\

3C 279 (FSRQ,LSP) & -0.02 & 7.95e-01 & 0.10 & 2.04e-01 & 0.24 & 1.03e-05 \\

PKS 1510-089 (FSRQ,LSP) & 0.13 & 3.18e-01 & -0.06 & 4.42e-01 & 0.09 & 1.54e-01 \\

PG 1553+113 (BLL,HSP) & -0.30 & 1.03e-01 & 0.09 & 4.19e-01 & 0.03 & 7.66e-01 \\

Mrk 501 (BLL,HSP) & -0.01 & 8.95e-01 & 0.07 & 4.06e-01 & -0.10 & 1.27e-01 \\

PKS 1749+096 (BLL,LSP) & -0.02 & 9.52e-01 & 0.09 & 6.35e-01 & -0.08 & 5.90e-01 \\

1ES 1959+650 (BLL,HSP) & -0.22 & 8.00e-02 & -0.41 & 5.04e-01 & -0.22 & 1.99e-02 \\

BL Lac (BLL,ISP) & -0.08 & 2.33e-01 & -0.03 & 7.34e-01 & -0.14 & 4.82e-03 \\

CTA 102 (FSRQ,LSP) & -0.20 & 9.48e-02 & 0.15 & 5.02e-02 & -0.11 & 7.41e-02 \\

3C 454.3 (FSRQ,LSP) & -0.18 & 2.88e-02 & -0.05 & 5.63e-01 & -0.35 & $<$ 1e-06 \\

1ES 2344+514 (BLL,HSP) & 0.13 & 4.65e-01 & 0.07 & 6.33e-01 & -0.12 & 2.47e-01 \\

\end{tabular}
\label{tab:spearman_results}
\end{table*}

\begin{table*}
\centering
\caption{Spearman correlation results for source classes.}
\begin{tabular}{lcccccc}
\hline
Type & \multicolumn{2}{c}{Active states} & \multicolumn{2}{c}{Non-active states} & \multicolumn{2}{c}{All states} \\
\hline
 &  $r$ & $p$ & $r$ & $p$ & $r$ & $p$ \\
\hline
BLL & 0.07 & 1.60e-02 & 0.06 & 5.89e-02 & -0.01 &  6.88e-01 \\

FSRQ & -0.16 & 8.94e-04 & -0.21 & 4.87e-08 & -0.19 & $<$ 1e-06 \\

LSP & 0.10 & 1.394e-02 & -0.05 &  1.46e-01 & 0.02 &  $<$ 1e-06\\

ISP & 0.07 & 8.24e-02 & 0.02 &  6.21e-01 & -0.04 & 1.99e-01 \\

HSP & -0.10 &1.32e-01 & 0.002 & 9.68e-01 & -0.12 & 6.42e-03 \\
\hline
\end{tabular}
\tablefoot{
For each source we report the Spearman rank correlation
coefficient ($r$) and the corresponding p-value ($p$) for active,
non-active, and combined states.
}

\label{tab:spearman_groups}
\end{table*}

\section{Results from the polarization angle rotations analysis}

\begin{onecolumn}

\begin{longtable}{lccccccc}
\caption{Optical polarization angle rotations.}\\
\hline\hline
Source & Amplitude & Start  & End & $\Delta{MJD}$  & p-t-test & p-binom & Rate \\
 & (deg) &  (MJD) &  (MJD) & (days) &  &  & (deg/day)\\
\hline
\endfirsthead
\caption{continued.}\\
\hline\hline
Source & Amplitude & Start  & End & $\Delta{MJD}$  & p-t-test & p-binom & Rate \\
 & (deg) &  (MJD) &  (MJD) & (days) &  &  & (deg/day)\\
\hline
\endhead
\hline
\endfoot

3C~66A & 98.0 $\pm$ 1.6 & 57177.44 & 57190.94 & 13.5 $\pm$ 0.67 & $6\times10^{-5}$ & 0.0625 & 7.26 $\pm$ 0.38 \\
AO~0235+164 & 152.9 $\pm$ 1.2 & 54844.7 & 54882.64 & 37.94 $\pm$ 3.15 & 0.0001 & 0.0078 & 4.03 $\pm$ 0.34 \\
AO~0235+164 & 118.0 $\pm$ 1.1 & 55094.91 & 55127.82 & 32.91 $\pm$ 5.63 & 0.0007 & 0.0312 & 3.59 $\pm$ 0.61 \\
AO~0235+164 & 129.1 $\pm$ 1.8 & 55127.82 & 55152.82 & 25.0 $\pm$ 5.63 & 0.0235 & 0.0625 & 5.16 $\pm$ 1.17 \\
AO~0235+164 & 140.7 $\pm$ 1.0 & 56899.06 & 56921.92 & 22.86 $\pm$ 4.75 & 0.0005 & 0.0312 & 6.15 $\pm$ 1.28 \\
AO~0235+164 & 90.2 $\pm$ 0.7 & 56921.92 & 56936.72 & 14.8 $\pm$ 4.75 & $2\times10^{-6}$ & 0.0312 & 6.09 $\pm$ 1.96 \\
AO~0235+164 & 158.2 $\pm$ 3.2 & 57430.65 & 57464.14 & 33.49 $\pm$ 2.07 & $1\times10^{-6}$ & 0.0078 & 4.72 $\pm$ 0.31 \\
AO~0235+164 & 109.7 $\pm$ 2.7 & 57694.84 & 57735.78 & 40.93 $\pm$ 3.91 & $2\times10^{-6}$ & 0.0625 & 2.68 $\pm$ 0.26 \\
AO~0235+164 & 400.3 $\pm$ 1.0 & 58060.33 & 58148.65 & 88.33 $\pm$ 0.46 & $6\times10^{-8}$ & 0.0004 & 4.53 $\pm$ 0.03 \\
S5~0716+714 & 126.7 $\pm$ 0.1 & 55594.76 & 55599.72 & 4.96 $\pm$ 0.14 & 0.0107 & 0.0625 & 25.53 $\pm$ 0.73 \\
S5~0716+714 & 103.8 $\pm$ 0.1 & 55976.79 & 56009.7 & 32.91 $\pm$ 3.86 & 0.0038 & 0.0625 & 3.15 $\pm$ 0.37 \\
S5~0716+714 & 145.1 $\pm$ 2.7 & 56423.15 & 56434.54 & 11.39 $\pm$ 3.22 & 0.0107 & 0.0312 & 12.74 $\pm$ 3.61 \\
OJ~287 & 123.8 $\pm$ 2.1 & 56578.12 & 56588.6 & 10.48 $\pm$ 1.51 & $7\times10^{-7}$ & 0.0078 & 11.81 $\pm$ 1.72 \\
OJ~287 & 151.9 $\pm$ 0.3 & 57014.89 & 57046.82 & 31.94 $\pm$ 0.91 & $7\times10^{-5}$ & 0.0078 & 4.76 $\pm$ 0.14 \\
W~Com & 100.7 $\pm$ 1.0 & 57130.35 & 57175.03 & 44.68 $\pm$ 2.16 & 0.0092 & 0.0625 & 2.25 $\pm$ 0.11 \\
3C~279 & 153.3 $\pm$ 1.9 & 55184.99 & 55245.89 & 60.9 $\pm$ 3.23 & $3\times10^{-10}$ & 0.0001 & 2.52 $\pm$ 0.14 \\
3C~279 & 102.2 $\pm$ 1.7 & 55292.8 & 55297.34 & 4.54 $\pm$ 3.23 & $6\times10^{-6}$ & 0.0625 & 22.53 $\pm$ 16.05 \\
3C~279 & 116.1 $\pm$ 0.9 & 55638.87 & 55659.34 & 20.46 $\pm$ 2.31 & $1\times10^{-7}$ & 0.0156 & 5.67 $\pm$ 0.64 \\
3C~279 & 104.9 $\pm$ 0.1 & 55710.73 & 55740.71 & 29.99 $\pm$ 1.5 & $1\times10^{-7}$ & 0.0156 & 3.5 $\pm$ 0.18 \\
3C~279 & 93.2 $\pm$ 0.4 & 55890.53 & 55948.99 & 58.45 $\pm$ 0.76 & $6\times10^{-16}$ & 0.0002 & 1.59 $\pm$ 0.02 \\
3C~279 & 137.2 $\pm$ 0.3 & 55971.94 & 56040.89 & 68.95 $\pm$ 0.76 & $9\times10^{-14}$ & 0.0001 & 1.99 $\pm$ 0.02 \\
3C~279 & 98.1 $\pm$ 0.6 & 56095.74 & 56126.18 & 30.44 $\pm$ 0.76 & $1\times10^{-5}$ & 0.0625 & 3.22 $\pm$ 0.08 \\
PKS~1510-089 & 122.8 $\pm$ 0.9 & 54948.88 & 54952.89 & 4.01 $\pm$ 2.26 & 0.0009 & 0.0312 & 30.65 $\pm$ 17.29 \\
PKS~1510-089 & 180.7 $\pm$ 8.6 & 55183.03 & 55227.01 & 43.98 $\pm$ 1.91 & 0.0004 & 0.0039 & 4.11 $\pm$ 0.27 \\
PKS~1510-089 & 169.8 $\pm$ 3.6 & 55922.54 & 55954.04 & 31.5 $\pm$ 4.93 & 0.0004 & 0.0039 & 5.39 $\pm$ 0.85 \\
PKS~1510-089 & 177.5 $\pm$ 1.5 & 56779.65 & 56798.82 & 19.17 $\pm$ 1.56 & 0.0008 & 0.0156 & 9.26 $\pm$ 0.76 \\
PKS~1510-089 & 137.2 $\pm$ 4.9 & 56879.78 & 56921.62 & 41.84 $\pm$ 1.56 & 0.0194 & 0.0625 & 3.28 $\pm$ 0.17 \\
PKS~1510-089 & 165.9 $\pm$ 4.9 & 57529.6 & 57596.7 & 67.1 $\pm$ 5.21 & 0.0002 & 0.0156 & 2.47 $\pm$ 0.21 \\
PKS~1510-089 & 239.4 $\pm$ 5.1 & 57598.51 & 57607.78 & 9.28 $\pm$ 5.21 & 0.0001 & 0.0156 & 25.81 $\pm$ 14.49 \\
PKS~1510-089 & 210.7 $\pm$ 8.4 & 57749.54 & 57781.04 & 31.5 $\pm$ 5.83 & 0.0001 & 0.0625 & 6.69 $\pm$ 1.27 \\
PKS~1510-089 & 317.3 $\pm$ 2.4 & 57823.91 & 57935.55 & 111.64 $\pm$ 5.83 & $3\times10^{-16}$ & $3\times10^{-6}$ & 2.84 $\pm$ 0.15 \\
PG~1553+113 & 104.1 $\pm$ 4.7 & 56720.27 & 56766.37 & 46.1 $\pm$ 5.77 & $2\times10^{-9}$ & 0.0039 & 2.26 $\pm$ 0.3 \\
PG~1553+113 & 142.1 $\pm$ 4.7 & 56776.5 & 56815.24 & 38.74 $\pm$ 5.77 & $2\times10^{-17}$ & $1\times10^{-5}$ & 3.67 $\pm$ 0.56 \\
PG~1553+113 & 150.3 $\pm$ 5.2 & 56856.67 & 56883.79 & 27.12 $\pm$ 5.77 & $1\times10^{-10}$ & 0.0009 & 5.54 $\pm$ 1.19 \\
PG~1553+113 & 197.1 $\pm$ 3.0 & 57587.83 & 57620.27 & 32.44 $\pm$ 2.33 & $2\times10^{-19}$ & $4\times10^{-7}$ & 6.08 $\pm$ 0.45 \\
PKS~1749+096 & 359.7 $\pm$ 2.8 & 56860.92 & 56900.34 & 39.42 $\pm$ 3.31 & $4\times10^{-6}$ & 0.0004 & 9.13 $\pm$ 0.77 \\
PKS~1749+096 & 134.8 $\pm$ 3.1 & 56901.85 & 56946.21 & 44.37 $\pm$ 3.31 & 0.0001 & 0.0625 & 3.04 $\pm$ 0.24 \\
PKS~1749+096 & 220.1 $\pm$ 1.9 & 57206.9 & 57245.34 & 38.43 $\pm$ 1.61 & $1\times10^{-6}$ & 0.0156 & 5.73 $\pm$ 0.24 \\
BL~Lac & 203.2 $\pm$ 1.9 & 56552.04 & 56572.01 & 19.97 $\pm$ 0.86 & $1\times10^{-8}$ & 0.0001 & 10.18 $\pm$ 0.45 \\
BL~Lac & 131.5 $\pm$ 2.0 & 56926.73 & 56951.55 & 24.82 $\pm$ 0.22 & 0.0001 & 0.0625 & 5.3 $\pm$ 0.09 \\
BL~Lac & 117.0 $\pm$ 1.6 & 57341.95 & 57356.47 & 14.52 $\pm$ 0.67 & $1\times10^{-5}$ & 0.0625 & 8.06 $\pm$ 0.39 \\
BL~Lac & 169.6 $\pm$ 0.5 & 57460.02 & 57473.02 & 12.99 $\pm$ 0.49 & $9\times10^{-5}$ & 0.0625 & 13.05 $\pm$ 0.5 \\
BL~Lac & 104.3 $\pm$ 3.2 & 57595.63 & 57604.94 & 9.32 $\pm$ 1.55 & $3\times10^{-10}$ & 0.0019 & 11.2 $\pm$ 1.9 \\
BL~Lac & 90.3 $\pm$ 0.2 & 57655.7 & 57687.69 & 31.99 $\pm$ 1.55 & $2\times10^{-5}$ & 0.0625 & 2.82 $\pm$ 0.14 \\
BL~Lac & 145.4 $\pm$ 0.4 & 58077.68 & 58096.6 & 18.92 $\pm$ 0.22 & $8\times10^{-6}$ & 0.0312 & 7.69 $\pm$ 0.09 \\
CTA~102 & 174.1 $\pm$ 3.4 & 56577.68 & 56616.94 & 39.25 $\pm$ 3.93 & $3\times10^{-9}$ & 0.0019 & 4.44 $\pm$ 0.45 \\
CTA~102 & 102.8 $\pm$ 4.6 & 56628.28 & 56652.63 & 24.35 $\pm$ 3.93 & $7\times10^{-6}$ & 0.0625 & 4.22 $\pm$ 0.71 \\
CTA~102 & 220.3 $\pm$ 2.8 & 56922.72 & 56952.09 & 29.38 $\pm$ 2.7 & $2\times10^{-8}$ & 0.0019 & 7.5 $\pm$ 0.7 \\
CTA~102 & 154.1 $\pm$ 2.4 & 56952.73 & 56957.65 & 4.92 $\pm$ 2.7 & 0.0009 & 0.0625 & 31.3 $\pm$ 17.2 \\
CTA~102 & 105.1 $\pm$ 1.3 & 57187.44 & 57223.4 & 35.96 $\pm$ 5.1 & $7\times10^{-5}$ & 0.0625 & 2.92 $\pm$ 0.42 \\
CTA~102 & 213.8 $\pm$ 5.8 & 57224.87 & 57271.97 & 47.1 $\pm$ 5.1 & 0.0004 & 0.0625 & 4.54 $\pm$ 0.51 \\
CTA~102 & 154.7 $\pm$ 3.4 & 57307.69 & 57336.01 & 28.32 $\pm$ 5.1 & $2\times10^{-8}$ & 0.0078 & 5.46 $\pm$ 0.99 \\
CTA~102 & 110.2 $\pm$ 9.0 & 57337.98 & 57347.73 & 9.75 $\pm$ 5.1 & $3\times10^{-6}$ & 0.0625 & 11.3 $\pm$ 5.98 \\
CTA~102 & 131.5 $\pm$ 4.0 & 57580.41 & 57598.67 & 18.26 $\pm$ 1.39 & $2\times10^{-6}$ & 0.0625 & 7.2 $\pm$ 0.59 \\
CTA~102 & 121.0 $\pm$ 4.5 & 57598.67 & 57612.52 & 13.85 $\pm$ 1.39 & $3\times10^{-14}$ & 0.0019 & 8.74 $\pm$ 0.94 \\
CTA~102 & 95.9 $\pm$ 2.8 & 57633.17 & 57636.1 & 2.93 $\pm$ 1.39 & $2\times10^{-6}$ & 0.0625 & 32.76 $\pm$ 15.57 \\
CTA~102 & 106.4 $\pm$ 0.6 & 57764.59 & 57781.59 & 16.99 $\pm$ 1.39 & $3\times10^{-15}$ & 0.0009 & 6.26 $\pm$ 0.51 \\
CTA~102 & 160.3 $\pm$ 1.5 & 58252.93 & 58282.88 & 29.95 $\pm$ 1.43 & $1\times10^{-5}$ & 0.0625 & 5.35 $\pm$ 0.26 \\
CTA~102 & 113.9 $\pm$ 1.1 & 58299.4 & 58310.85 & 11.45 $\pm$ 1.43 & $1\times10^{-6}$ & 0.0312 & 9.95 $\pm$ 1.25 \\
3C~454.3 & 178.8 $\pm$ 5.3 & 56494.55 & 56540.06 & 45.5 $\pm$ 6.26 & $3\times10^{-6}$ & 0.0002 & 3.93 $\pm$ 0.55 \\
3C~454.3 & 161.3 $\pm$ 4.7 & 56869.56 & 56885.52 & 15.95 $\pm$ 1.1 & $7\times10^{-5}$ & 0.0019 & 10.11 $\pm$ 0.75 \\
3C~454.3 & 108.4 $\pm$ 8.7 & 56902.97 & 56917.01 & 14.04 $\pm$ 1.1 & 0.0452 & 0.0625 & 7.72 $\pm$ 0.87 \\
3C~454.3 & 100.3 $\pm$ 2.4 & 57336.5 & 57343.51 & 7.01 $\pm$ 1.22 & 0.0001 & 0.0312 & 14.3 $\pm$ 2.52 \\
3C~454.3 & 167.9 $\pm$ 4.2 & 57605.53 & 57612.58 & 7.04 $\pm$ 1.12 & 0.0002 & 0.0312 & 23.83 $\pm$ 3.84 \\
\hline

\label{tab:rotations}
\end{longtable}
\tablefoot{Columns list the source name, rotation amplitude,
start and end dates in MJD, duration, t-test p-value, binomial p-value,
and rotation rate.}
\end{onecolumn}

\end{appendix}

\end{document}